\documentclass[11pt,showpacs]{revtex4}
\input epsf.sty

\usepackage{ulem}
\usepackage{cancel}
\usepackage{color}

\def\comment#1{}

\usepackage{graphicx}
\begin{document}
\title{Do high-energy neutrinos travel faster than photons in a discrete space-time?}
\author{She-Sheng Xue}
\email{xue@icra.it}
\affiliation{ICRANeT, Piazzale della Repubblica, 10-65122, Pescara,\\
Physics Department, University of Rome ``La Sapienza'', Rome,
Italy} 


\begin{abstract}
The recent OPERA measurement of high-energy neutrino velocity, once independently verified, implies new physics in the neutrino sector. We revisit the theoretical inconsistency of the fundamental high-energy cutoff attributing to quantum gravity with the parity-violating gauge symmetry of local quantum field theory describing neutrinos. This inconsistency suggests high-dimension operators of neutrino interactions.  Based on these studies, 
we try to view the OPERA result, high-energy neutrino oscillations and indicate to observe the restoration of parity conservation by measuring the asymmetry of high-energy neutrinos colliding with left- and right-handed polarized electrons.    
  
\end{abstract}

\maketitle

\noindent
{\bf Introduction.}
\hskip0.1cm
Since their appearance, neutrinos have always been extremely peculiar. Their
charge neutrality, near masslessness, flavour mixing, oscillation and parity-violating coupling have been at the center of a conceptual elaboration and
an intensive experimental analysis that have played a major role in
donating to mankind the beauty of the standard model for particle physics.  
Recently, the OPERA experiment \cite{exp} reports  anomaly in flight time of neutrinos ($\langle E_\nu\rangle = 17$ GeV) from CERN to INFN Gran Sasso. This anomaly corresponds to a relative
difference of the muon neutrino velocity with respect to the speed of light $(v-c)/c = (2.48 \pm 0.28 \,({\rm stat.})\pm 0.30\, ({\rm sys.})) 10^{-5}$. As described in Ref.~\cite{exp},  
this value is by far much larger than the relative
deviation from the speed of light c of the neutrino velocity due to its finite rest mass, that is expected to be smaller than $10^{-19}$. In the past, a high energy ($E_\nu > 30$ GeV) and short baseline experiment has been able to
test deviations down to $|v-c|/c < 4\times 10^{-5}$ \cite{expnv1}. With a baseline analogous to that of OPERA but at
lower neutrino energies ($E_\nu$ peaking at $\sim 3$ GeV with a tail extending above $100$ GeV), the MINOS
experiment reported a measurement of $(v-c)/c = 5.1 \pm 2.9\times 10^{-5}$ \cite{expnv2}. A larger deviation of the neutrino velocity from c would be a striking result pointing to new physics in the neutrino sector.
Given the potential far-reaching consequences of such a result, independent measurements are needed before the effect can either be refuted or firmly established. 
Nevertheless, in this letter, 
we revisit the theoretical inconsistency of the fundamental high-energy cutoff attributing to quantum gravity with the parity-violating gauge symmetry of local quantum field theory describing neutrinos. This inconsistency suggests high-dimension operators of neutrino interactions.  Based on these studies, 
we try to view the OPERA result, high-energy neutrino oscillations and indicate to observe the restoration of parity conservation by measuring the asymmetry of high-energy neutrinos colliding with left- and right-handed polarized electrons.

In continuum space-time manifold without any length scale at short distances, the Lorentz symmetry group is unbounded at the high boost (or high
energy) end. Exact Lorentz symmetry may be broken in the high boost
limit \cite{coleman}, if there is a fundamental cutoff, the Planck length (scale) $a_{\rm pl}\sim 10^{-33}\,$cm ($\Lambda_{\rm pl}=\pi/a_{\rm pl}\sim 10^{19}\,$GeV) at short distances of space-time, intrinsically attributing to quantum gravity.
The possibility of Lorentz symmetry breaking  has been considered in many sophisticate theories, 
Loop quantum gravity \cite{lqg1}, string theory and non-commutative geometry
\cite{string1,ncom1}, standard
model extension with
Lorentz symmetry breaking operators \cite{kas}, and phenomenological quantum gravity model  \cite{qgliv,piran}.

We may conceive that precisely due to the violent fluctuations that the gravitational field {\it must}
exhibit at $a_{\rm pl}$, space-time ``ends'' there. Either by the creation of a Wheeler ``foam'' \cite{wheeler1964}, or by some other mechanism which we need not discuss here, one may conceive that as a result the physical space-time gets endowed with a fundamental Planck length, $a_{\rm pl}$, and thus the basic arena of physical reality becomes a {\it random lattice with lattice constant} $a_{\rm min}\sim a_{\rm pl}$ \cite{preparata91}.  We recently calculate this minimal length $a_{\rm min}\approx 1.2\,a_{\rm pl}$ \cite{xue2010} in 
studying quantum Einstein-Cartan theory in the framework of Regge calculus and its variant \cite{wheeler1964,regge61,hammer_book}. 

This discrete space-time provides a natural regulator for local quantum field theories for particles and gauge interactions. A natural regularized quantum field theory demands the existence of ultra-violet fix points where renormalization group invariance follows so that low-energy observables are independent of high-energy cutoff. Based on low-energy observations of parity violation, the Lagrangian of standard model was built preserving exact chiral-gauge symmetries $SU_L(2)\otimes U_Y(1)$ that are accommodated by left-handed lepton doublet $(\nu_i, i)_L$ and right-handed single $(i)_R$ ($i=e,\mu,\tau$). On the other hand, a profound result
obtained 30 years ago, in the form of a no-go theorem \cite{nn1981}, tells us that there is no any consistent way to transpose straightforwardly on a discrete space-time the bilinear neutrino Lagrangian of the continuum theory exactly preserving chiral gauge symmetries. This no-go theorem was also demonstrated \cite{nn1991} to be generic and independent of discretization (regularization) scheme of space-time. Either one gived up the gauge principle by 
explicitly breaking chiral gauge symmetries \cite{wilson} or we were led to consider at least quadrilinear (four) neutrino interactions to preserve chiral-gauge symmetries \cite{xue_chiral}.  This is not the place for a detailed discussion of this important result and its possible relation to quantum gravity. We focus ourselves to discussions that in discrete space-time, Lorentz symmetry is broken, bilinear neutrino kinetic terms do not preserve chiral-gauge symmetries, and quadrilinear (four) neutrino interactions are introduced in connection with recent on going high-energy neutrino experiments. (The natural units $\hbar=c=1$ are adopted, unless otherwise specified).

\noindent
{\bf Energy-momentum relations.}
\hskip0.1cm In order to simplify discussions, we start with a hypercube lattice with lattice constant $a$, 
instead of a complex discrete space-time like the simplicial manifold discussed for quantum gravity \cite{xue2010}.
On a spatial lattice (time continuum), the Klein-Gorden motion for a free massive gauge boson is \cite{kogut}
\begin{eqnarray}
\ddot \phi = \frac{1}{a^2}\Delta \phi(x) -m^2c^4\phi(x),
\label{rep}
\end{eqnarray}
where the differentiating operator
\begin{eqnarray}
\Delta \phi(x) \equiv \sum_\mu\left[\phi(x+a\hat n_\mu) +\phi(x-a\hat n_\mu) -2 \phi(x)\right],
\label{delta}
\end{eqnarray}
and the energy-momentum relation is given by 
\begin{eqnarray}
E^2=m^2+\frac{2}{a^2}w(k),\quad  w(k)\equiv 2\sin^2(\frac{ka}{2})
\label{kg3}
\end{eqnarray}
and 
\begin{eqnarray}
E^2\approx m^2+k^{\,2}-\frac{1}{12}(k^4a^2)+\cdot\cdot\cdot\, , \label{kg4}
\end{eqnarray}
for $ka\ll 1$. Therefore, the velocity of a massless boson (photon) is given by 
\begin{eqnarray}
v_\gamma = \frac{dE_\gamma}{dk}\approx 1-\frac{1}{8}(ka)^2\approx 1-\frac{1}{8}(E_\gamma a)^2. 
\label{gv}
\end{eqnarray}
which is smaller than the speed of light traveling 
in a continuum space-time. 

As discussed in the introductory paragraph, chiral symmetries have to be explicitly broken if 
Hamiltonian $\mathcal H$ is bilinear in left-handed (Weyl) neutrino fields $\psi$, 
\begin{eqnarray}
{\mathcal H}&=&\sum_x\Big\{-\frac{i}{2a}\bar\psi(x)\gamma_\mu [\psi(x+a\hat n_\mu)-\psi(x-a\hat n_\mu)]\nonumber\\
&-&\frac{B}{2a}\bar\psi(x)\Delta\psi_R(x) +m_\nu \bar\psi(x)\psi_R(x)\Big\},
\label{fermioh}
\end{eqnarray}
where $\psi_R$ is a right-handed neutrino, and the second term is an explicit chiral-symmetry-breaking term \cite{wilson}. 
The energy-momentum relation is \cite{kogut} 
\begin{eqnarray}
E_\nu^2=m_\nu^2+\frac{\sin^2ka}{a^2}+B^2\frac{w^2(k)}{a^2},
\label{drf3}
\end{eqnarray}
and
\begin{eqnarray}
E_\nu^2 &\simeq & k^2+m_\nu^2+\frac{1}{4}\left(B^2-\frac{4}{3}\right)k^4a^2+\cdot\cdot\cdot,
\label{drf4}
\end{eqnarray}
for $ka\ll 1$.
Analogously, the velocity of neutrinos is 
\begin{eqnarray}
v_\nu &=& \frac{dE_\nu}{dk}
\approx  1-\frac{1}{2}\frac{m_\nu^2}{E_\nu^2}+\frac{3}{8}\left(B^2-\frac{4}{3}\right)(E_\nu a)^2, 
\label{nv}
\end{eqnarray}
where $k^2\approx E^2_\nu\gg m^2_\nu $.
For the Wilson parameter $B^2>4/3$ and large neutrino energy $E_\nu$, Eq.~(\ref{nv}) implies the possibility that high-energy neutrinos can travel faster than photons 
in discrete space-time. 
In this low-energy region, photon and neutrino energies $E_{\gamma,\nu}$ are much smaller than the cutoff scale $\Lambda=\pi/a$ of discrete space-time ($E_{\gamma,\nu}\ll \Lambda$),  
the Lorentz symmetry is restored, energy-momentum relations $E_\gamma=k$ and $E_\nu \approx k\gg m_\nu$, velocities $v_\gamma = 1$ and $v_\nu < 1$.

Assuming a more complex discrete space-time, inspired by Eqs.~(\ref{kg4},\ref{drf4}) one may parameterize energy-momentum relations as follow
\begin{eqnarray}
E^2_\gamma &=&
k^2[1 - (k/\Lambda_\gamma)^{\alpha_\gamma}],\label{pgkg}\\
\quad E^2_\nu &=& m^2_\nu+
k^2[1 + (k/\Lambda_\nu)^{\alpha_\nu}] 
\label{gkg}
\end{eqnarray}
where parameters are indexes $\alpha_{\gamma,\nu}=1,2,\cdot\cdot\cdot$, and the characteristic scales $\Lambda_{\gamma,\nu}$ of Lorentz symmetry breaking.
Eqs.~(\ref{pgkg}) and (\ref{gkg}) are reminiscent of  phenomenological energy-momentum relations adopted for studying arrival time variations of high-energy cosmic particles due to Lorentz symmetry breaking \cite{piran}. 
Suppose that $\alpha_\nu= 2$, $\Lambda_{\nu,\gamma}\sim\Lambda_{\rm pl}\simeq 10^{19}\,$GeV the Planck scale due to quantum gravity and neutrino energy (mass) $E_\nu\approx 17\,$GeV ($m_\nu\sim 1\,$eV),  neutrino velocity-variation is negligibly small, as given by Eq.~(\ref{nv}), because of Planck-scale suppression.  The second term due to neutrino mass $-(m_\nu/E_\nu)^2\sim -10^{-20}$ and last term due to discrete space-time gives $+10^{-36}$, as a result the neutrino velocity-variation $(v_\nu-c)/c\approx -10^{-20}+ 10^{-36}\lesssim 0$, and the neutrino velocity is smaller than the speed of light ($v_\nu\lesssim c$).
Analogously, Eq.~(\ref{gv}) gives very small velocity-variation of photons $(v_\gamma-c)/c\approx -10^{-36}\lesssim 0$ due to discrete space-time. 
In order to observe such small effect of Lorentz symmetry breaking
due to quantum gravity, one is bound to detect variations of arrival times of high-energy gamma ray and neutrinos from astrophysical sources at cosmological distances.

\begin{figure}
\begin{center}
\includegraphics[width=2.5in]{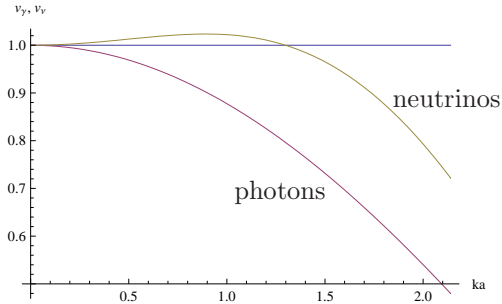}
\put(-95,40){photons}
\put(-35,75){neutrinos}
\caption{The photon and neutrino velocities in a discrete space-time are plotted as functions of $ka$, and $k_{\rm max}a=1.3$ for for $B=(3/2)^{1/2}=1.225$.}
\label{velocity1}
\end{center}
\end{figure}

Instead of small-$ka$ expansions (\ref{drf4}) and (\ref{nv}), using Eqs.~(\ref{kg3}) and (\ref{drf3}), we calculate photon  velocity $dE_\gamma/dk$ and neutrino velocity $dE_\nu/dk$, which are plotted 
in Fig.~\ref{velocity1}. We find that (i) the photon velocity $v_\gamma < 1$ for any value of $ka$; (ii)  the neutrino velocity $v_\nu > 1$ for $k\in [k_{\rm min},k_{\rm max}]$; (iii) 
$v_\nu < 1$ for $k\notin [k_{\rm min},k_{\rm max}]$. We approximately obtain
\begin{eqnarray}
k_{\rm min}a \approx  (am_\nu)^{1/2}(2^{3/4}/B^{1/2});\quad
k_{\rm max}a \approx  2\arcsin (1+2B/10)^{1/2}.
\label{kminmax}
\end{eqnarray}
\comment{
Numerical calculations show a critical value $B_c\approx 1.164$, above which $B>B_c$, it appears the energy-range $[k_{\rm min}, k_{\rm max}]$ where $v_\nu > 1$. This energy-range $[k_{\rm min}, k_{\rm max}]$ increases as $B$-value  increases.
} 

The recent OPERA result shows $(v_\nu -c)/c\approx 2\times 10^{-5}$ for average neutrino energy $\langle E_\nu \rangle\approx 17\,$GeV \cite{exp}, and this neutrino velocity anomaly roughly occurs in the energy range ($15$GeV--$45$GeV). 
Using Eq.~(\ref{gkg}), we find $\Lambda_\nu\sim  10^3\,$GeV, which is much smaller than the Planck scale $\Lambda_{\rm pl}$.
This is indeed surprising, because we do not have any compelling reason to introduce at this energy scale either an explicit chiral-symmetry breaking or a spontaneous chiral-symmetry 
breaking with new Goldstone bosons for violations of chiral gauge and Lorentz symmetries. 
This implies that the OPERA result would be very important to indicate new physics and this is also one of reasons to call further confirmations of the OPERA result. Nevertheless, we start to revisit a theoretical scenario, 15 years old, to view the OPERA result, 
high-energy neutrino oscillations and indicate to observe the restoration of parity conservation by measuring the asymmetry of high-energy neutrinos colliding with left- and right-handed polarized electrons.

\noindent
{\bf Four-neutrino interactions.}
\hskip0.1cm
As discussed in the introductory paragraph, in order to preserve chiral symmetries of Lagrangian in a discrete space-time, we introduced a chiral-symmetric interaction of four neutrino fields \cite{xue_chiral}
\begin{eqnarray}
\frac{g}{4a^2}\sum_{x,i} \bar\psi^i_L(x)\Delta\psi_R(x)\Delta\bar\psi_R(x)\psi^i_L(x),
\label{4f}
\end{eqnarray}
where $SU_L(2)$ doublets $\psi^i_L(x)=(\nu_i, i)$ carry lepton flavors indicated by ``$i=e,\mu, \tau$'', $\psi_R(x)$ a unique gauge singlet, and strong four-neutrino coupling $g\gg 1$.  The Ward identity of the shift-symmetry $\psi_R(x)\rightarrow \psi_R(x) + {\rm const.}$ leads to the one-particle irreducible vertex of four-neutrino interacting, 
\begin{eqnarray}
\Gamma^{(4)}=g\,w(p+q/2)\,w(p'+q/2),
\label{4fe}
\end{eqnarray}
where $p+q/2$ and $p'+q/2$ are the momenta of the $\psi_R$ field; $p-q/2$ and $p'-q/2$ are the momenta of the $\psi_L^i$ field ($q$ is the momentum transfer). In high energies, namely large neutrino energy and energy transfer, $\Gamma^{(4)}$ is so large that bound states of neutrinos,  $\Psi^i_R\sim (\bar\psi_R\psi^i_L)\psi_R$, are form and carry opposite chirality of left-handed neutrinos $\psi^i_L$. Namely, $\Psi^i_R$ is a right-handed $SU_L(2)$ doublet.
Left-handed neutrino $\psi^i_L$ and right-handed composite neutrino $\Psi^i_R$ couple to intermediate gauge bosons in the same way, as a consequence the parity symmetry is restored. In addition, $\psi^i_L$ combines with $\Psi^i_R$ to form a Dirac neutrino $\Psi^i_D=(\psi^i_L, \Psi^i_R)$, whose inverse propagator obtained by the method of strong coupling expansion in terms of $(1/g)$ 
\cite{xue_chiral, xue2008}
\begin{eqnarray}
S^{-1}_{ij}(k)= \delta_{ij}\frac{i}{a}\gamma_\mu \sin(k_\mu a)+ \delta_{ij}M(k),
\label{infp}
\end{eqnarray}
where we neglect small neutrino masses $m_\nu$ and 
\begin{eqnarray}
M(k)=\frac{g}{a}\,w(k),
\label{inmass}
\end{eqnarray}
is chiral-gauge-symmetric masses of composite Dirac neutrinos \cite{neutral}. In the light of Occam's razor, we assume that four-neutrino coupling $g$ is independent of lepton flavors so that the spectrum (\ref{infp}) is diagonal in the flavor space. Instead, normal neutrino mass ($m_\nu$) matrix is not diagonal.
 
In low energies, namely small neutrino energy and energy transfer, $\Gamma^{(4)}$ becomes so small that the binding energies $E_{\rm bind}(g,a)$ of bound states of three neutrinos 
vanishes, as a result bound states dissolve into their constituents \cite{pole-cut} and the mass term $M(k)$ (\ref{inmass}) vanishes, that we call dissolving phenomenon. Therefore,
neutrino family contains three left-handed neutrinos $\nu_e,\nu_\mu,\nu_\tau$ and the parity symmetry is violated as described by the standard model \cite{c-symmetry}, in addition to a right-handed neutrino $\psi_R=\nu_R$ \cite{r-neutrino}.
 
However, to quantitatively show such dissolving phenomenon in low-energies by using Eqs.~(\ref{infp}) and (\ref{inmass}) obtained in high-energies, one is bound to find a ultra-violet fix point and renormalization group equation in the neighborhood of ultra-violet fix point, where the dimension-10 interaction (\ref{4f}) receives anomalous dimensions and becomes renormalizable dimension-4 operator. 
This is a complicate and difficult issue and needs non-perturbative calculations.     
Nevertheless, in Refs.~\cite{xue_chiral,xue_parity}, we assumed a characteristic energy threshold ``${\mathcal E}$'' at which the dissolving phenomenon occurs 
and postulated that the energy threshold ${\mathcal E}$ be much larger than electroweak symmetry breaking scale $\Lambda_{\rm ew}\sim 250\,$GeV, 
\begin{equation}
\Lambda_{\rm ew}\ll {\mathcal E} < \Lambda_{\rm pl}.
\label{epsilon}
\end{equation} 
The energy threshold ${\mathcal E}={\mathcal E}(g,a)$ also characterizes the energy scale of Lorentz symmetry breaking due to non-vanishing mass term $M(k)$ (\ref{inmass}).
In this letter, in order to obtain an effective energy-momentum relation in low-energy, 
we further assume that as the four-neutrino coupling $g(a)$ is running with $a$ the lattice constant, an effective four-neutrino coupling $G_X(g,a)\equiv a{\mathcal E}g(a)$ 
follows renormalization group equation in the neighborhood of ultra-violet fix point. 
In consequence, Eq.~(\ref{infp}) gives the energy-momentum relation
\begin{eqnarray}
E_\nu^2=m_\nu^2+\frac{\sin^2ka}{a^2}+\frac{(2G_X)^2}{a^2}\sin^4(ka/2),
\label{idrf3}
\end{eqnarray}
and for $ka\ll 1$
\begin{eqnarray}
E_\nu^2 &\simeq & k^2+m_\nu^2+[(2G_X)^2-\frac{1}{3}]k^4a^2+\cdot\cdot\cdot,
\label{idrf4p}
\end{eqnarray}
where we replace the four-neutrino coupling $g$ by the effective one $G_X$. From Eq.~(\ref{idrf4p}) and $(2G_X)^2\gg 1/3$, the velocity of high-energy neutrinos is given by
\begin{eqnarray}
v_\nu &=& \frac{dE_\nu}{dk}\approx 1 - \frac{1}{2}\left(\frac{m_\nu}{E_\nu}\right)^2 + (2G_X)^2(ka)^2+\cdot\cdot\cdot\nonumber\\
&\approx & 1 -\frac{1}{2}\left(\frac{m_\nu}{E_\nu}\right)^2 +\left(\frac{E_\nu}{\bar{\mathcal E}}\right)^2+\cdot\cdot\cdot\,, 
\label{inv}
\end{eqnarray}
where $\bar{\mathcal E}={\mathcal E}/g$, $k^2\gg m^2_\nu$ and $k\approx E_\nu$.
Eq.~(\ref{inv}) implies that high-energy neutrino velocity $v_\nu>1$,  
provided the composite Dirac neutrino (\ref{infp}) is formed, the energy-threshold and neutrino energy obey $2^{1/2}E_\nu/\bar{\mathcal E}> m_\nu/E_\nu$.
\comment{
Using the OPERA result $(v_\nu -c)/c\approx 2\times 10^{-5}$ for average neutrino energy $E_\nu \approx 17\,$GeV \cite{exp}, we find that Eq.~(\ref{inv}) gives ${\mathcal E} \sim  10^3\,$GeV \cite{confirm}.
} 
Low-energy neutrino velocity $v_\nu<1$, for the reason that the composite Dirac neutrino (\ref{infp}) is not formed, the mass term $M(k)$ (\ref{inmass}) vanishes and the third term in Eq.~(\ref{inv}) is absent. As well known, at much lower energy, in the
$10$ MeV range, a stringent limit of $|v-c|/c < 2\times 10^{-9}$ was set by the observation of (anti) neutrinos emitted by the SN1987A supernova \cite{expnv3}. 

\begin{figure}
\begin{center}
\includegraphics[width=2.5in]{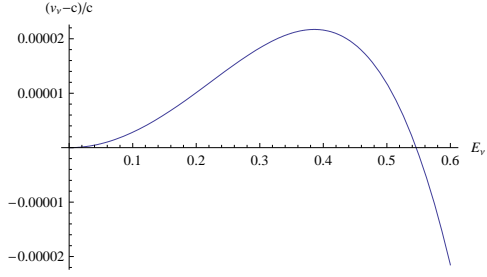}
\caption{Based on Eq. (\ref{idrf4}),  the neutrino velocity $v_\nu=dE_\nu/dk$ is numerically calculated for $(2G_X)^2=5.35$ and
the neutrino velocity variation $(v_\nu-c)/c$ is plotted as a function of the 
neutrino energy $E_\nu\approx k_\nu$ in unit of the energy-scale $\bar\mathcal E=\mathcal E/g$.}
\label{vvx}
\end{center}
\end{figure}

In unit of the energy-scale $\bar {\mathcal E}$,
we rewrite the energy-momentum relation (\ref{idrf3}),
\begin{eqnarray}
E_\nu^2=m_\nu^2+(2G_X)^2\sin^2(k/2G_X)+(2G_X)^4\sin^4(k/4G_X), 
\label{idrf4}
\end{eqnarray}
and the chiral symmetric masses (\ref{inmass})
\begin{eqnarray}
M(k)=(2G_X)^2\sin^2(k/4G_X).
\label{inmassr}
\end{eqnarray} 
Using Eq. (\ref{idrf4}), we numerically calculate 
the neutrino velocity $v_\nu=dE_\nu/dk$ and  plot the neutrino velocity-variation $(v_\nu-c)/c$ in Fig.~\ref{vvx}. 
Numerical calculations show a critical value $G_X^{\rm crit}\approx 1.15$ [$(2G_X^{\rm crit})^2=5.335$], above which $G_X>G_X^{\rm crit}$, it appears an intermediate energy-range $[E^{\rm min}_\nu, E^{\rm max}_\nu]$ where neutrino velocity-variation $(v_\nu-c)/c>0$. This energy-range and velocity-variation 
depend on three parameters $G_X$, $\mathcal E$ and $g$. 
Current high-energy neutrino experiments might start to gain an insight into this theoretical scenario. 
\comment{
, and neutrino velocity-variation
can be consistent with the OPERA result for appropriate values of two parameters
This strongly implies that $G_c$ could be a ultra-violet fix point for low-energy physics, $G(g,a)\rightarrow G_c$ and $aG(g,a)\rightarrow (\mathcal E)^{-1}$, as the running coupling $g(a)$ follows the renormalization invariant group equation in the neighborhood of ultra-violet fix point, where $\xi/a\rightarrow \infty$, where $\xi$ is the correlation length. from theoretical point of view, all these need non-perturbative calculations, however, we can try to use $G>G_c$ as a parameter to see whether or not the solution given by this scenario can be consistent with experiment result and fix the energy-threshold $\mathcal E\sim 10^{3}$GeV, which is actually characteristic scale of the problem.
In this stage, we do not attempt to do a fit with experimental data by adjusting the parameter $G$, instead, we just want to show our scenario is self-consistency and not inconsistent with experimental result, since both theoretical scenario and experimental results need to be further improved.  It is possible to get into the range of the OPERA experiment result
for the coupling $G$ approaching to the cortical value $G_c$, as shown in Fig.~(\ref{vvx}). 
The invariant mass (\ref{inmassr}) $M(k)\approx (E_\nu/4{\mathcal E})^2{\mathcal E}$, which is at least ${\mathcal E}$ for 
$E_\nu\gtrsim {\mathcal E}$, and $M(k)>{\mathcal E}$ for $E_\nu >{\mathcal E}$.
}

We turn to the discussions of high-energy neutrino oscillations in this scenario. For a two-flavor system, 
considering
Eq.~(\ref{infp}) and normal neutrino masses $m_\nu$, the Hamiltonian in the base of
flavor eigenstates is given by
\begin{eqnarray}
{\mathcal H}_{\rm flavor} &=& E_\nu+\frac{m_1^2+m_2^2}{4E_\nu}+\left(\frac{\Delta
m^2_{12}}{4E_\nu}\right)
\left(%
\begin{array}{cc}
  -\cos2\theta & \sin2\theta \\
  \sin2\theta & \cos2\theta \\
\end{array}%
\right)+\left(%
\begin{array}{cc}
  M & 0 \\
 0 & M \\
\end{array}%
\right), 
\label{h-flavor}
\end{eqnarray}
where $\Delta m^2_{12}=m^2_2-m^2_1$, the first part is a unitary transformation of
the Hamiltonian in the base of mass eigenstates 
\begin{eqnarray}
\mathcal H_{\rm mass}&=&\left(%
\begin{array}{cc}
  E_1 & 0 \\
  0 & E_2 \\
\end{array}%
\right)\simeq E_\nu+\left(%
\begin{array}{cc}
  m^2_1/2E_\nu & 0 \\
  0 & m^2_2/2E_\nu \\
\end{array}%
\right)
\label{hmass},
\end{eqnarray}
and the leading contribution to neutrino energy $E_{1,2}$ is obtained by 
assuming $k_1\approx k_2\approx E_\nu$, and $k_{1,2}a\ll 0$; 
the unitary transformation $U$ and mixing angle $\theta$ are given by
\begin{eqnarray}
U=\left(%
\begin{array}{cc}
  \cos\theta & \sin\theta \\
  -\sin\theta & \cos\theta \\
\end{array}%
\right),\quad
\tan 2\theta&=&\frac{2\hat{\mathcal H}_{12}}{\hat{\mathcal
H}_{22}-\hat{\mathcal H}_{11}}. 
\label{theta}
\end{eqnarray}
Note that that four-neutrino coupling $g$ and chiral symmetric mass $M$ (\ref{inmass}) do not depends on lepton flavors, otherwise, the missing angle $\theta$ would depend on neutrino energy \cite{Kostelecky}.
As a result, the energy-difference between neutrino flavors $E^i_\nu-E^j_\nu= \Delta m^2_{12}/E_\nu$ and the probability of two flavor neutrino oscillation,
\begin{eqnarray}
P_{i\leftrightarrow j}(L)&\approx&\sin^22\theta\sin^2\left[(E^i_\nu-E^j_\nu)L \right],\label{pro2}
\end{eqnarray}
where  $L$ distance neutrino travels.

In addition, the parity symmetry should be restored when the composite Dirac neutrino (\ref{infp}) is formed above 
the energy-threshold $({\mathcal E})$, as already mentioned.
For example,
\comment{
because the discussions are straightforward as long as chiral symmetries are exactly preserved
For the reason that the three-fermion states (\ref{threeb}) carry the $SU_L(2)$
quantum number, there must be an interacting 1PI vertex between $W^\pm,Z^\circ$
bosons
and composite right-handed fermions (\ref{threeb}) in the high-energy region.
}
the effective one-particle irreducible interacting vertex between neutrino ($p$), electron ($p'$) and $W^\pm$ gauge boson ($q=p'-p$)  \cite{xue_parity}, 
\begin{eqnarray}
\Gamma^{ij}_\mu (p,p')&=&i\frac{g_2}{2\sqrt{2}}U_{ij}\gamma_\mu \left[P_L+f(p,p')\right]
\label{wv}
\end{eqnarray}
where $g_2$ is the $SU_L(2)$ coupling, $U_{ij}$ the CKM matrix, and
left-handed projector $P_L=(1-\gamma_5)/2$. The vector-like (parity conserving) form factor $f(p,p')$ is related to the chiral-symmetric mass (\ref{inmass}) by the Ward identity of chiral gauge symmetries. In high-energies, $f(p,p')\not=0$ for $p,p'\ge{\mathcal E}$, and chiral-gauge coupling becomes vector-like. Similar discussions and calculations for chiral-gauge coupling to the neutral gauge boson $Z^0$ can be found in Ref.~\cite{xue_parity}. In consequence, the parity symmetry is restored. 
This could be experimentally checked by measuring the left-right asymmetry
\begin{eqnarray}
A_{LR}&=&\frac{\sigma_L-\sigma_R}{\sigma_L+\sigma_R}
\label{aslr}
\end{eqnarray}
where $\sigma_L$ ($\sigma_R$) is the cross-section of high-energy neutrino colliding with left-handed (right-handed) polarized electron. We expect the restoration of parity symmetry, $A_{LR}\rightarrow 0$, 
for high-energy neutrinos and electrons. 
In low-energies,  bound states of neutrinos dissolve into their constituents, the vector-like form factor $f(p,p')$ in Eq.~(\ref{wv}), 
\begin{equation}
f(p,p')|_{p,p'\rightarrow{\mathcal E}+0^+}\rightarrow 0,\quad
f(p,p')|_{p,p'<{\mathcal E}}= 0,
\label{threshold}
\end{equation}
which leads to the parity-violating gauge-coupling ($P_L$-term in Eq.~(\ref{wv})), and $A_{LR}=1$ as measured in low-energy experiments.
All these features could possibly be verified or falsified by current high-energy neutrinos experiments.

\noindent
{\bf Some remarks.}
\hskip0.1cm 
We end this letter by making some remarks. Due to quantum gravity, the very-small-scale structure of space-time and high-dimensional operators of fermion-interactions must be much more complex \cite{torsion}. Because it is difficult to have a complete theory at the Planck scale and carry out non-perturbative calculations leading to observables at low-energy scale. In our theoretical scenario, we adopt the simplest space-time discretization and four-neutrino interaction (\ref{4f}), leading to a phenomenological model in low-energies.
The recent experiment for high-energy neutrino \cite{exp}, if it is further confirmed, might begin to shed light on this most elusive and fascinating arena of fundamental particle physics. 

In a recent paper \cite{Giudice}, based on the radiative correction to an electron propagator via exchanging a W-boson within the standard model, authors show
\begin{equation}
\delta v_e = g_{\rm su2}^2\int\frac{d^4k}{ (4\pi)^4}\frac{ \delta v_\nu(k)}{ (k^2)[(k+p)^2-M_W^2]}\gtrsim
\left(\frac{E_{\rm OPERA}}{ (4\pi)\Lambda_{\rm ew}}\right)^2\delta c_\nu(E_{\rm OPERA})\sim 10^{-9},
\label{constrain}
\end{equation}
for the OPERA result $\delta v_\nu(E_{\rm OPERA})\sim 10^{-5}$ at $E_{\rm OPERA}\sim 20$GeV. Eq.~(\ref{constrain}) is clearly inconsistent with the experimental bound on electron velocity-variation $\delta v_e< 10^{-14}$. This is indeed a problem that implies two possibilities either the OPERA result is wrong or the standard model needs some modifications. In the scenario we present, high-energy neutrino acquires a large chiral-symmetric mass $M$ (\ref{inmass}) and gauge-coupling to W-bosons is no longer purely left-handed (\ref{wv}). Therefore, (i) the radiative correction to an electron propagator contains not only kinetic term, but also mass term, (ii) neutrino propagator $\sim 1/(k^2+M^2)$, 
this can give a suppression factor 
$(k/M)^2$ in the right-handed side of Eq.~(\ref{constrain}). It looks a possibility to get around this problem. In future work, one needs  
\comment{
that the OPERA result could still possibly be consistent with the observational limit of electron velocity variation. In this scenario, the radiative corrections to the electron propagator should be calculated by using the neutrino propagator (\ref) and W-boson propagator (\ref), and $Z_1=Z_2$ due to exact chiral-gauge symmetries, where $Z_1$ is the renormalized function for electron propagetor and $Z_2$ the renormalized function for the vertex coupling to $W^\pm$-gauge bosons.
} 
to do some detailed calculations of radiative corrections to the propagators of neutrinos and charged leptons in the scenario presented in this letter. 

In another recent paper \cite{Glashow}, based on the standard model, authors discuss the Cherenkov radiation of superluminal neutrinos via the process 
$\nu\rightarrow \nu +e^++e^-$ with the energy-threshold $E_0=2m_e/(v_\nu^2-v_e^2)^{1/2}$. The OPERA result $\delta v_\nu\approx 10^{-5}$ and $\delta v_e< 10^{-14}$ lead to $E_0=140$ MeV. This indicates that it is impossible to observe $17$--$45$ GeV superluminal neutrinos, 
because of energy-lost via Cherenkov radiation. 
This again implies that either the OPERA result is wrong or the standard model needs some modifications. In this scenario, in addition to the four-fermion interaction (\ref{4f}) for the set of left-handed $SU_L(2)$-doublets, there is an analogous four-fermion interaction for right-handed $SU_L(2)$-singlets \cite{xue_chiral,xue_parity},
\begin{eqnarray}
\frac{g}{4a^2}\sum_{x,i} \bar\psi^i_R(x)\Delta\nu_L(x)\Delta\bar\nu_L(x)\psi^i_R(x),
\label{4fr}
\end{eqnarray}
where singlet $\psi^i_R=e_R,\mu_R,\tau_R$ and $\nu_L$ is the left-handed counterpart of $\nu_R$ \cite{r-neutrino}.
Analogously, strong coupling ($g\gg 1$)  
leads to form left-handed bound states (singlets), 
for example $(\bar\nu_L\cdot e_R)\nu_L$, which combines with $e_R$ to form a composite Dirac particle $[e_R, (\bar\nu_L\cdot e_R)\nu_L]$
with a large chiral-gauge invariant mass $M$ (\ref{inmass}). In this case, electron mass $m_e$ 
should be replaced by the mass $M$ in the energy-threshold $E_0=2M/(v_\nu^2-v_e^2)^{1/2}$, which becomes much larger than $140$MeV for $M\gg m_e$. This may provide a possibility that the OPERA result is not in contradiction to energy-lost via Cherenkov radiation. In future work, some detailed analysis of Cherenkov radiation of superluminal neutrinos should be carried out in the scenario presented in this letter. We would like to mention some early papers on superluminal neutrinos as tachyons 
\cite{tachyon}.
At the present situation, we are facing two possibilities: (i) the OPERA result is confirmed to be wrong; (ii) the OPERA result is confirmed to be correct and further neutrino experiments should be proposed to indicate what modifications standard model needs at high-energies.

\noindent
{\bf Acknowledgment.}
\hskip0.1cm
The author thanks M.~Creutz, G.~t'Hooft, L.~Maiani, H.B.~Nielson, G.~Preparata and M.~Testa for the discussions on lattice chiral fermions, H.~Kleinert for the discussions on neutrino physics, torsion and strong coupling expansion. The author thanks G.~Amelino-Camdedia for reading the manucript and communications on the issue, and Roberto Battiston for clear explanations on the OPERA result. The author also thanks G.~Preparata and R.~Ruffini who led his attention to Wheeler's foam \cite{wheeler1964}, Planck-scale structure of space-time, the arena of physical reality. Finally, the 
author thanks anonymous referee for his/her questions and comments that greatly improve the manuscript.


\begin{thebibliography}{99}
\bibitem{pole-cut}
Bound states of three neutrinos are presented as poles of four-neutrino interacting $S$-matrix, locating at $E_{\rm bind}(g,a)<0$ in the energy-momentum plane. When the coupling $g$ becomes smaller, $E_{\rm bind}\rightarrow 0^-$, these poles of the four-neutrino interacting $S$-matrix become cuts, representing three elementary neutrino states.
\bibitem{c-symmetry} In the phase transition of neutrino bound states $\Psi_R\sim (\bar\psi_R\psi)\psi_R$ formation (dissolution) from (into) their constitutes, chiral symmetries are exactly preserved, and realized by different chiral fermion (neutrino) contents \cite{xue_chiral}.  It
should be pointed out that this phenomenon is not a spontaneous symmetry
breaking and no Goldstone bosons are produced \cite{njl}.
\bibitem{r-neutrino}
In Ref.~\cite{xue_chiral, xue_parity}, it can be found that the Ward identity of the shift-symmetry $\nu_{R,L}(x)\rightarrow \nu_{R,L}(x) + {\rm const.}$ requires the right(left)-handed neutrino $\nu_{R,L}$ to be free particles that could be candidates of warm dark matter.
\end{thebibliography}
\end{document}